\documentclass{pnastwomodified}
\usepackage{amssymb,amsfonts,amsmath}
\usepackage{graphicx}

\begin{document}

%%%%%%%%%%%%%%%%%%%%%%%%%%%%%%

%% For titles, only capitalize the first letter

\title{Finding and testing network communities by lumped Markov chains}

\author{Carlo Piccardi\affil{1}{Department of Electronics and Information, Politecnico di Milano, Piazza Leonardo da Vinci 32, I-20133 Milano, Italy, email: carlo.piccardi@polimi.it}}

%\contributor{Submitted to Proceedings of the National Academy of Sciences
%of the United States of America}

\maketitle

%%%%%%%%%%%%%%%%%%%%%%%%%%%%%%%%%%%%%%%%%%%%%%%%%%%%%%%%%%%%%%%%
\begin{article}

\begin{abstract}
Identifying communities (or clusters), namely groups of nodes with comparatively strong internal connectivity, is a fundamental task for deeply understanding the structure and function of a network. Yet, there is a lack of formal criteria for defining communities and for testing their significance. We propose a sharp definition which is based on a significance threshold. By means of a lumped Markov chain model of a random walker, a quality measure called ``persistence probability'' is associated to a cluster. Then the cluster is defined as an ``$\alpha$-community'' if such a probability is not smaller than $\alpha$. Consistently, a partition composed of $\alpha$-communities is an ``$\alpha$-partition''. These definitions turn out to be very effective for finding and testing communities. If a set of candidate partitions is available, setting the desired $\alpha$-level allows one to immediately select the $\alpha$-partition with the finest decomposition. Simultaneously, the persistence probabilities quantify the significance of each single community. Given its ability in individually assessing the quality of each cluster, this approach can also disclose single well-defined communities even in networks which overall do not possess a definite clusterized structure.
\end{abstract}


%% The first letter of the article should be drop cap: \dropcap{}
%\dropcap{I}n this article we study the evolution of ''almost-sharp'' fronts

%% Enter the text of your article beginning here and ending before
%% \begin{acknowledgements}
%% Section head commands for your reference:
%% \section{}
%% \subsection{}
%% \subsubsection{}

\dropcap{C}omplex networks are currently one of the most extensively studied subjects in the field of applied mathematics. In the last fifteen years, a huge number of theoretical results have been put forward, and almost any field of science and technology has benefit from the application of such results to specific problems \cite{St:01,BoLa:06,BaBa:08,Ne:10}.

One of the most promising but challenging tasks in network science is \textit{community analysis}, which is aimed at revealing possible partitions of a network into subsets of nodes (\textit{communities}, or \textit{clusters}) with dense intra- but sparse inter-group connections.  Finding and analyzing such partitions often provides invaluable help in deeply understanding the structure and function of a network, as widely demonstrated by several case studies in social sciences \cite{Ne:06,GuSa:07}, biology \cite{JoCa:06}, economics \cite{PiCa:10}, or information science \cite{FlLa:02}, just to name a few.

Despite the abundance of contributions on this subject (see \cite{Fo:10} for a survey), the issue of community analysis cannot be considered satisfactorily solved. First of all, finding communities is a computationally hard task, because the ``best'' partition must be sought for in a set whose cardinality grows faster than exponentially with the number of nodes. The exhaustive enumeration of the partitions is thus impossible, and heuristic techniques must be employed. Secondly, and perhaps more important, there is no widespread consensus on formal criteria for defining communities and for testing their significance \cite{Fo:10}. When a subnetwork can actually be considered to form a community, namely a group of nodes with comparatively strong internal connectivity? Probably the most important attempt to answer this question was put forward by Newman and coworkers \cite{NeGi:04,Ne:06,Ne:06b}, who defined a quality index called \textit{modularity} which quantifies, for a given partition of the network into candidate communities, to what extent the distribution of the intra-/inter-community edges is anomalous with respect to a suitably defined random network. Since high modularity values are obtained in presence of groups of nodes with comparatively large intra-community edge density, maximizing modularity should put in evidence the ``best'' partition. This method has been proven successful in many circumstances but, on the other hand, it has been widely demonstrated that, due to intrinsic limitations, it does not necessarily always yield a significant partition \cite{ReBo:06,FoBa:07,Fo:10}. And even when it does, it quantifies the quality of a partition but not of each individual community.

This paper introduces a sharp definition of community which is based on a threshold of significance. More precisely, once a level $0<\alpha<1$ is specified, a node cluster is defined to be an $\alpha$\textit{-community} if the probability that a random walker, which is currently in one of the cluster's nodes, remains in the cluster in the next step is not smaller than $\alpha$. Such a probability is obtained from an approximate \textit{lumped Markov chain} model of the random walker (i.e., a reduced-order Markov chain in which the communities of the original network become nodes) which is easily derived from the original (high-order) Markov chain model. Consistently, a partition composed of $\alpha$-communities is defined to be an $\alpha$\textit{-partition}.

If equipped with an effective method for generating a set of ``good'' candidate partitions, the notions of $\alpha$-community and $\alpha$-partition provide a framework for simultaneously finding communities and testing their significance. For that, the desired significance level $\alpha$ is first fixed. Then, a family of partitions is derived and each partition is immediately checked to assess whether it is formed by $\alpha$-communities. This allows one to identify the $\alpha$-partitions, and to select one of them. Typically, one searches for communities which are at the same time small (to effectively decompose the network) and significant (with much more internal than external connectivity). For that, a guideline is that of selecting, among the available $\alpha$-partitions, the one with the largest number of communities.

But the notion of $\alpha$-community can also be useful in a partially different way. It may happen that, for a given significance level $\alpha$, no $\alpha$-partitions are found. Yet, one or a few $\alpha$-communities could exist. They correspond to strongly connected groups of nodes, even in a network which, overall, does not possess a definite clusterized structure. Or, finally, one can assess the significance of the results of a single-partition method, such as modularity optimization \cite{Ne:06}, and obtain an immediate assessment of the $\alpha$-significance of each single community and, consequently, of the entire partition.

In the paper, we first introduce the lumped Markov chain model of the random walker and define the notions of $\alpha$-community and $\alpha$-partition. Testing the $\alpha$-significance of a given community or partition turns out to be extremely parsimonious in computational terms. Then we analyze a few examples of application and, for that, we propose an effective algorithm for deriving a meaningful set of partitions. The algorithm, which applies hierarchical cluster analysis, is again based on the Markov chain model of a random walker and, consequently, it involves a notion of similarity/distance among nodes which is consistent with the significance criterion above introduced. We finally compare this approach, which can be applied to fully general networks (i.e., directed and weighted), with other community analysis methods having a similar philosophy.

%%%%%%%%%%%%%%%%%%%%%%%%%%%%%%%%%%%%%%%%%%%%%%%%%%%%%%%

\section{Networks, $\alpha$-Communities, and $\alpha$-Partitions }\label{sec:net}

Consider a network with nodes $\mathbb{N}=\{1,2,\ldots,N\}$ and $L$ edges. We consider the most general case of directed and weighted network, and we denote by
$W=[w_{ij}]$ the $N\times N$ weight matrix, where
$w_{ij}\geq 0$ is the weight of the edge $i\rightarrow j$. The connectivity matrix $A=[a_{ij}]$ is
the $N\times N$ binary matrix where $a_{ij}=1$ if $w_{ij}>0$, and $a_{ij}=0$ otherwise. If the network is actually undirected we have $W=W'$ and $A=A'$, and if it is unweighted we let $W=A$ (i.e., all weights equal to $1$). We assume that the network is strongly connected (e.g., \cite{BaBa:08}), namely there exists an oriented path from any $i$ to any $j$. If the network is directed, for each node $i$ we define the (total) degree as $k_i=k^{in}_i+k^{out}_i=\sum_j a_{ji}+\sum_j a_{ij}$, whereas $k_i=\sum_j a_{ji}=\sum_j a_{ij}$ for undirected network. The average degree is given by $\langle k\rangle=\sum_i k_i/N$. Similarly, for a directed network the in-, out-, and total strength of node $i$ are given by $s^{in}_i=\sum_j
w_{ji}$, $s^{out}_i=\sum_j w_{ij}$, and $s_i=s^{in}_i+s^{out}_i$,
respectively, and the total network strength by
$s=\sum_{ij}w_{ij}$. If the network is undirected we have instead $s_i=s^{in}_i=s^{out}_i=\sum_j
w_{ji}=\sum_j w_{ij}$ and $s=\sum_{ij}w_{ij}/2$.

A $N$-state Markov chain $\pi_{t+1}=\pi_{t}P$, with $\pi_t= \left( \pi_{1,t} \pi_{2,t} \ldots
\pi_{N,t} \right)$, can be associated to the $N$-node network
by row-normalizing the weight matrix $W$, namely by letting the
transition probability from $i$ to $j$ equal to
    \begin{equation}\label{eq:p}
    p_{ij}=\frac{w_{ij}}{\sum_j w_{ij}}=\frac{w_{ij}}{s_i^{out}}.
    \end{equation}
The quantity $p_{ij}$ is the probability that a random walker which is in node $i$ jumps to node $j$, and $\pi_{i,t}$ is the probability of being in node $i$ at time $t$. The transition matrix $P=[p_{ij}]$ is a row-stochastic (or
Markov) matrix ($0\le p_{ij}\le 1$ for all $i,j$, and
$\sum_j p_{ij}=1$ for all $i$). Furthermore, $P$ is irreducible since the network is connected. This implies that the equation $\pi=\pi P$ has a unique solution $\pi$, which is strictly positive ($\pi_i>0$ for all $i$) \cite{Me:00} and corresponds to the stationary Markov chain state
probability distribution. For undirected networks one can easily check that $\pi=\left(s_1 s_2 \ldots s_N\right)/(2s)$, whereas for directed networks a general closed form does not exist and $\pi$ has to be numerically computed.

We denote by $\mathbb{P}_q$ a partition of $\mathbb{N}$ in $q$ subsets (or subnetworks), namely $\mathbb{P}_q=\{ \mathbb{C}_1,\mathbb{C}_2,\ldots,\mathbb{C}_q\}$ with $\bigcup_c \mathbb{C}_c=\mathbb{N}$ and $\mathbb{C}_c \cap
\mathbb{C}_d=\oslash$ for all $c,d$. In rough terms, the sub-network $\mathbb{C}_c$ is called a \textit{community} (or \textit{cluster}) if it has a high internal density of weight, i.e., if the total weight of the edges internal to $\mathbb{C}_c$ is much
larger than that of the edges connecting $\mathbb{C}_c$ to the
rest of the network. The community analysis of a given network consists therefore in finding the ``best''
partition $\mathbb{P}$, according to some criterion.
Despite a huge amount of contributions, there is however no widespread consensus on formal criteria for defining communities
and for testing their significance \cite{Fo:10}. As a consequence, in many situations a more fruitful approach is that of searching for a few, ``good'' partitions $\mathbb{P'},\mathbb{P''},\ldots$, among which selecting with common sense and experience.

Defining a partition $\mathbb{P}_q$ induces a $q$-state meta-network, where communities become meta-nodes. The rigorous description of the dynamics of the random walker at this scale by a \textit{lumped Markov chain}, however, is not possible if not in special cases \cite{KeSn:76} - actually, the Markovian property is not even preserved in general. Despite this limitation, a $q$-state Markov chain can be defined, which correctly describes the random walker at the aggregate level provided the stochastic process is started at the stationary distribution $\pi$ \cite{Bu:94,HoSa:09}. This lumped Markov chain is defined by the $q\times q$ row-stochastic matrix
    \begin{equation}\label{eq:U}
        U=\left[\mathrm{diag}\left(\pi H\right)\right]^{-1}H'\mathrm{diag}(\pi) PH,
    \end{equation}
where $H$ (\textit{collecting matrix}) is a $N\times q$ binary
matrix coding the partition $\mathbb{P}_q$, i.e., its entry $h_{ic}$ is $1$ if
and only if node $i\in\mathbb{C}_c$. The lumped Markov chain $\Pi_{t+1}=\Pi_{t}U$ shares the stationary distribution with the original one (suitably collected), namely $\Pi=\pi H$ satisfies $\Pi=\Pi U$. On the contrary, starting from an arbitrary $\pi_0$, the lumped Markov chain $\Pi=\Pi U$ started at $\Pi_0=\pi_0 H$ provides, in general, only an approximate description of the evolution of $\pi H$. The difference between the real and approximate $\Pi$, however, tends exponentially to zero if the two chains are regular \cite{Me:00}, since they converge, by definition, to the same stationary state.

The ability of the lumped Markov chain to describe the random walk dynamics only at stationarity is not a limitation for our purposes, as it will be demonstrated by the examples of application. Note that the entry $u_{cd}$ of $U$ is the probability that the random walker is at time $(t+1)$ in any of the nodes of community $d$, provided it is at time $t$ in any of the nodes of community $c$. The diagonal term $u_{cc}$ is defined \textit{persistence probability} of community $c$. Large values of $u_{cc}$ are expected for significant communities. In fact, the expected escape time from $\mathbb{C}_c$ is $\tau_c=(1-u_{cc})^{-1}$: the walker will spend long time within the same community if the weights of the internal edges are comparatively large with respect to those pointing outside. Given a value $0<\alpha<1$, $\mathbb{C}_c$ is defined $\alpha$-\textit{community} if $u_{cc}\geq\alpha$. Thus $\alpha$ acts as a selection parameter, as sharply qualifies communities with respect to a given threshold of significance. Consistently, $\mathbb{P}_q$ is defined $\alpha$-\textit{partition} if it is composed of $\alpha$-communities, namely $u_{cc}\geq\alpha$ for all $c=1,2,\ldots,q$.

\begin{figure*}
    \begin{center}
    \includegraphics[width=16cm]{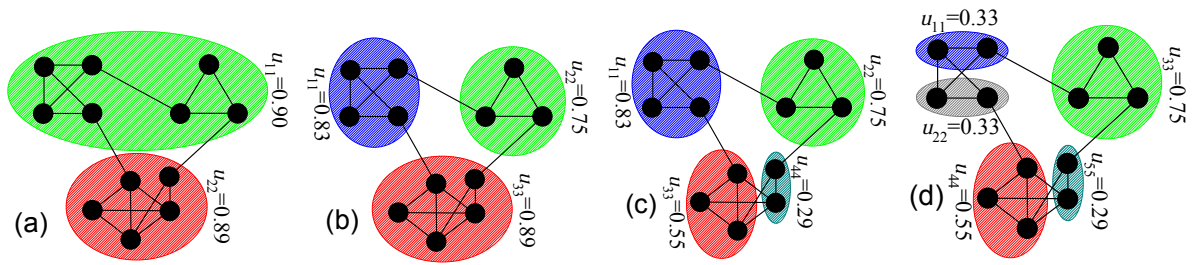}
    \caption{Four different partitions (with increasing number $q$ of communities) of the same network. The persistence probabilities $u_{cc}$ remain rather large as long as the network is partitioned into ``natural'' communities. Passing from (b) to (c), and from (c) to (d), significant communities are broken,with a sudden drop of the relevant persistence probabilities.}\label{fig:partitions}
    \end{center}
\end{figure*}

Consider the simple $12$-node network of Fig. \ref{fig:partitions} \cite{FoCa:09}, which is purposely composed of three clusters. Four partitions are considered, corresponding to finer and finer divisions, and the $u_{cc}$-s are computed for each candidate community. As long as the latter coincide with the ``natural'' communities, or with the union of two of them, all the $u_{cc}$-s are rather large. But as soon as a natural community is broken, some very low persistence probabilities are found. This result can be used in a twofold manner, as extensively shown in the next section. On one hand, if a set of finer and finer partitions is analyzed, the sudden drop of a $u_{cc}$ is the signal that a significant community has been broken: the network decomposition must be stopped before this event. On the other hand, if a single partition is given and its significance has to be assessed, the $u_{cc}$-s immediately quantify the quality of the partition but also of each individual community.

%%%%%%%%%%%%%%%%%%%%%%%%%%%%%%%%%%%%%%%%%%%%%%%%%%%%%%%

\section{\label{sec:examples}Applications and Examples}

The proposed method is now applied to assess the significance of sets of partitions related to a variety of networks. An algorithm for deriving partitions is first introduced, implementing hierarchical cluster analysis after a random-walk-based node distance is defined. Then the results related to three networks are discussed: a synthetical benchmark network with built-in cluster structure; a real-world network with a rather strong community structure; and another real-world network with weak clustering but with a few well-defined communities. 

\subsection{\label{sec:partitions}Deriving Partitions} Cluster analysis can be used to group ``similar nodes'' into candidate communities. This needs defining a meaningful \textit{similarity/distance} among each pair of nodes. Such a definition is by no means obvious: among the many proposals \cite{Fo:10}, a few exploit random walks to induce a suitable similarity measure (e.g., \cite{Zh:03,PoLa:05,FoPi:07,RoBe:08,StCh:10}). We follow this line by proposing an approach in which, however, we do not explicitly
perform random walks in a Monte Carlo fashion, but derive analytically the global behavior of a large number $M$ of walkers (a ``fleet'') started from each node $i$.

Consider a large number $M$ of repetitions of a
random walk started from $i$. For each repetition, the probability
that the walker is in $j$ after $t$ steps is $[P^t]_{ij}$. Thus,
if $M$ random walks of length $T$ are performed from $i$, the
expected number of visits to $j$ in any time instant in $1\leq t \leq T$ is
$M\sum_{t=1}^T [P^t]_{ij}$. By averaging with respect to $M$, we propose a (symmetric) similarity $\sigma_{ij}$ defined by
    \begin{equation}\label{eq:s}
    \sigma_{ij}=\sigma_{ji}=\sum_{t=1}^T \left( \left[ P^t \right]_{ij}+ \left[ P^t \right]_{ji}
    \right).
    \end{equation}
Note that this is conceptually
equivalent to an explicit random walk approach, but with an arbitrarily large number $M$ of
repetitions from each starting node instead of one only. Most notably, the results do not depend on the actual stochastic realization of the random walks. We finally define the distance $d_{ij}=d_{ji}$ between nodes $(i,j)$ by complementing the similarity and normalizing the
results between $0$ and $1$:
    \begin{equation}\label{eq:d}
    d_{ij}=d_{ji}=1-\frac{\sigma_{ij}-\min \sigma_{ij} }{\max \sigma_{ij} - \min     \sigma_{ij}}.
    \end{equation}
The rationale underlying the definition of $s$ and $d$ is to assign nodes $(i,j)$ a large similarity if a numerous fleet of random walkers started in $i$ (resp. $j$) makes a large number of visits to $j$ (resp. $i$) within a sufficiently small time horizon $T$. The notion of community induced by this metric, therefore, is that of a subnetwork where a random walker has a large probability of circulating for quite a long time, before eventually leaving to reach another group. This is conceptually consistent with the definition of $\alpha$-community above introduced. The choice of the time horizon $T$ is potentially critical: if too large, the
probability of visiting a given state $j$ becomes independent of the
starting state since it tends to $\pi_j$, whereas if $T$ is
too small the information gathered is insufficient. We will return later to this point.

\subsection{\label{sec:lfr}LFR benchmark} Lancichinetti, Fortunato, and Radicchi (LFR) \cite{LaFo:08} proposed a family of synthetically generated graphs, explicitly designed to serve as benchmarks for testing community detection algorithms. They explicitly take into account two properties found in real networks, namely the heterogeneity in the distributions of node degrees and community sizes. Both of the latter are taken as power laws, with prescribed exponents $\gamma$ and $\beta$, respectively. In addition, the network is defined by prescribing the number $N$ of nodes, the average degree $\langle k\rangle$, and a \textit{mixing parameter} $\mu$ such that each node shares a fraction $1-\mu$ of its edges with the other nodes of its own community, and a fraction $\mu$ with the rest of the network. The benchmark generating method was later extended to oriented and weighted networks \cite{LaFo:09b} - here we consider an example of an undirected, unweighted network with $N=1000$, $\langle k\rangle=20$, $\mu=0.25$, $\gamma=2$, and $\beta=1$. The network we obtained turns out to be formed by $38$ communities, with dimensions ranging from $10$ to $49$ nodes each.

\begin{figure}
    \begin{center}
    \includegraphics[width=8cm]{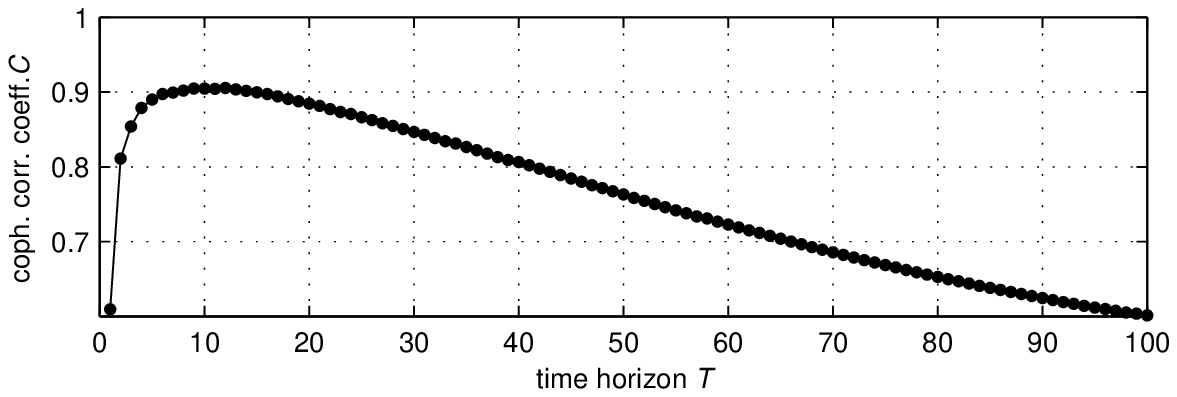}
    \includegraphics[width=8cm]{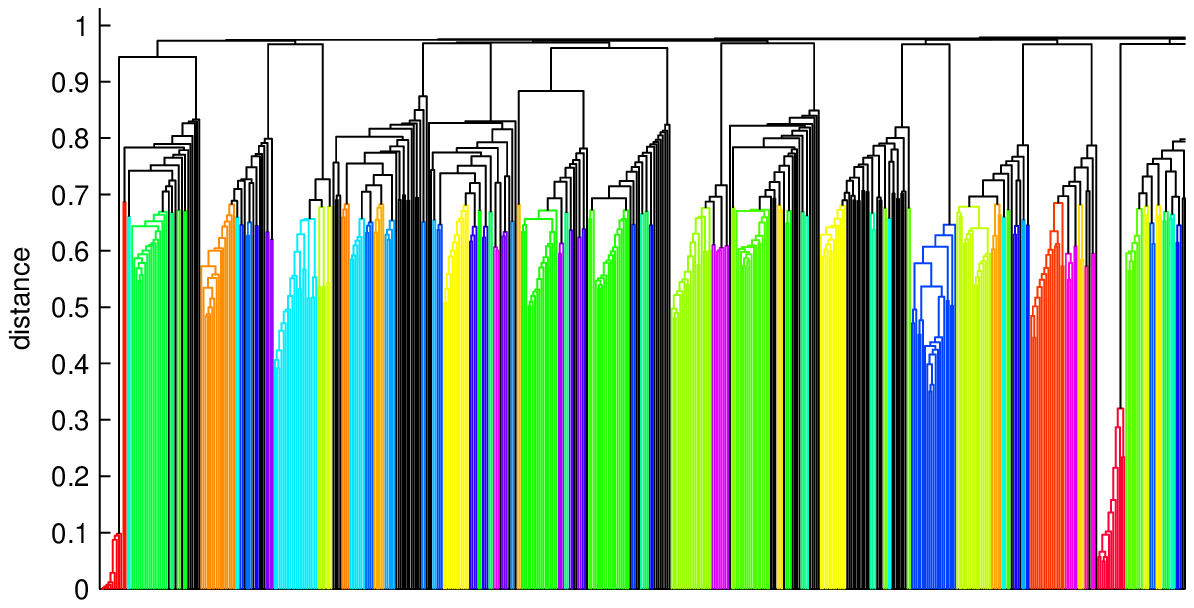}
    \caption{LFR benchmark network. Above: The cophenetic correlation coefficient $C$ as a function of $T$. The maximum is attained at $T=12$. Below: The dendrogram obtained with $T=12$
     (only half of the plot is presented for readability). } \label{fig:lfr}
    \end{center}
\end{figure}

Cluster analysis yields a different dendrogram for each time horizon $T$, whose choice is thus nontrivial. At the two extremes, setting $T=1$ restricts the pairs of nodes which are candidate to nonzero similarity to neighboring pairs only, whereas larger and larger values of $T$ tend to make any node equally similar to any other. We found that an effective selection of $T$ can be empirically obtained by maximizing the
\textit{cophenetic correlation coefficient} $C$, which is defined
as the linear correlation between the distances
${d_{ij}}$ and the \textit{cophenetic distances} ${c_{ij}}$
\cite{EvLa:11}. The latter are a product of the hierarchical
cluster analysis: for any node pair $(i,j)$, the cophenetic
distance $c_{ij}$ is the height of the link joining (directly or
indirectly) nodes $(i,j)$ in the dendrogram. The value of $C$ is generally used to assess whether the adopted distance $d_{ij}$ induces an effective clusterization (notice that $C$ qualifies the entire dendrogram, and not a network partition), although limitations have been observed in specific applications \cite{Ho:78}. Figure \ref{fig:lfr} shows the dependence of $C$ on $T$: we take $T=12$, which attains the maximum $C=0.905$. The related dendrogram is in the same figure.

\begin{figure}
    \begin{center}
    \includegraphics[width=8cm]{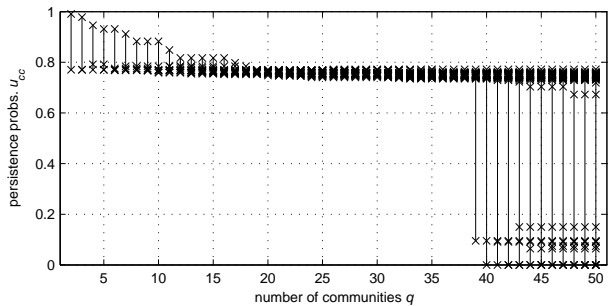}
    \caption{The persistence probabilities' diagram of the LFR benchmark network. For a partition with $q$ clusters, crosses denote the values of the $q$ diagonal terms $u_{cc}$ of the matrix $U$. Vertical straight lines are only for visual aid.} \label{fig:lfr_ucc}
    \end{center}
\end{figure}

Horizontal top-down cross-sections of the dendrogram identify a sequence $\mathbb{P}_2,\mathbb{P}_3,\ldots$ of partitions with increasing number of candidate communities. For each $\mathbb{P}_q$ we compute $U$ according to (\ref{eq:U}), and plot its diagonal terms in the \textit{persistence probabilities' diagram} of Fig. \ref{fig:lfr_ucc}. The diagram reveals a sharp discontinuity. For $q\leq 38$, all the $u_{cc}$-s are rather large ($u_{cc}\geq 0.735$ for all $c$). This means that significant communities are identified: in rigorous terms, all the proposed partitions $\mathbb{P}_q$ with $2\leq q\leq 38$ are $\alpha$-partitions with $\alpha=0.735$. For $q\geq 39$  significant communities are broken, as revealed by the sudden drop of a larger and larger number of $u_{cc}$-s. Remind that $38$ is exactly the number of communities planted in the synthetically generated network. It is worth mentioning that, if we search for the max-modularity partition (we used the so-called ``Louvain algorithm'' \cite{BlGu:08}, proved to be one of the most reliable \cite{LaFo:09}), we obtain a partition with $q=34$ communities, with modularity $Q=0.714$. The number of communities of the planted partition is thus not perfectly recovered (small communities tend to be aggregated). Nonetheless, for the obtained $\mathbb{P}_{34}$ partition the persistence probabilities are in the range $0.737 \leq u_{cc} \leq 0.772$, which is qualitatively consistent with the results of Fig. \ref{fig:lfr_ucc}.

We have finally compared the built-in partition planted in the LFR benchmark network with the partition $\mathbb{P}_{38}$ obtained with our method, as well as with the ``max-modularity'' partition. The comparison is in terms of the \textit{normalized mutual information} $I$, a reliable and often used measure of partition similarity, introduced by  \cite{DaDi:05} to the network research community. Here we only point out that $I=1$ when the two partitions are identical, whereas $I$ has zero expected value for independent partitions. We obtained $I=0.992$ for the partition resulting from our method (actually, we checked that as few as $0.08\%$ of the pairs $i,j$ are misclassified), and a slightly smaller $I=0.987$ for the ``max-modularity'' partition.

\subsection{\label{sec:netscience}Netscience network}

The Netscience network is a weighted,
undirected, social network describing the collaborations (up to year 2006)
among researchers in network science, the weight of the edge
connecting two researchers being proportional to the number of papers
they have co-authored \cite{Ne:06b}. Its giant component has
$N=379$ nodes, and it is generally considered an example of a
real network with a rather strong community structure. Many methods for network analysis, included community detection algorithms, have been tested and discussed on this example (e.g., \cite{DeYa:10,NaNa:11,CaHa:11}).

\begin{figure}
    \begin{center}
    \includegraphics[width=8cm]{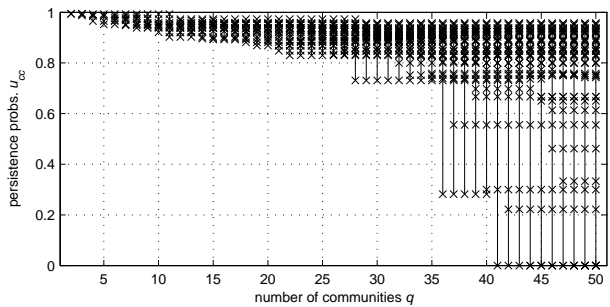}
    \caption{The persistence probabilities' diagram of the Netscience network. } \label{fig:netscience_ucc}
    \end{center}
\end{figure}

At $T=6$ we get the dendrogram with largest $C$, and the resulting persistence probabilities' diagram is in Fig. \ref{fig:netscience_ucc}. The plot has a less clear structure than that of the LFR network (Fig. \ref{fig:lfr_ucc}): the proper $q$ must be selected with a trade-off between a finer decomposition (large $q$) and a higher significance of the communities (small $q$). For example, all the partitions with $q$ up to $10$ are $\alpha$-partitions with $\alpha>0.9$. But, if less stringent significance levels are required, partition with $q\leq 27$, or even $q\leq 35$, seem to be perfectly meaningful.

\begin{figure}
    \begin{center}
    \includegraphics[width=8cm]{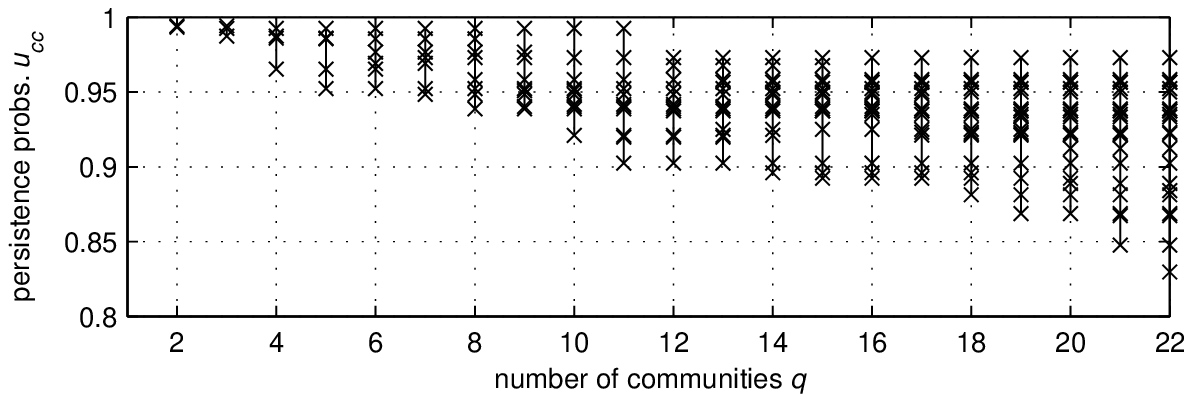}
    \includegraphics[width=8cm]{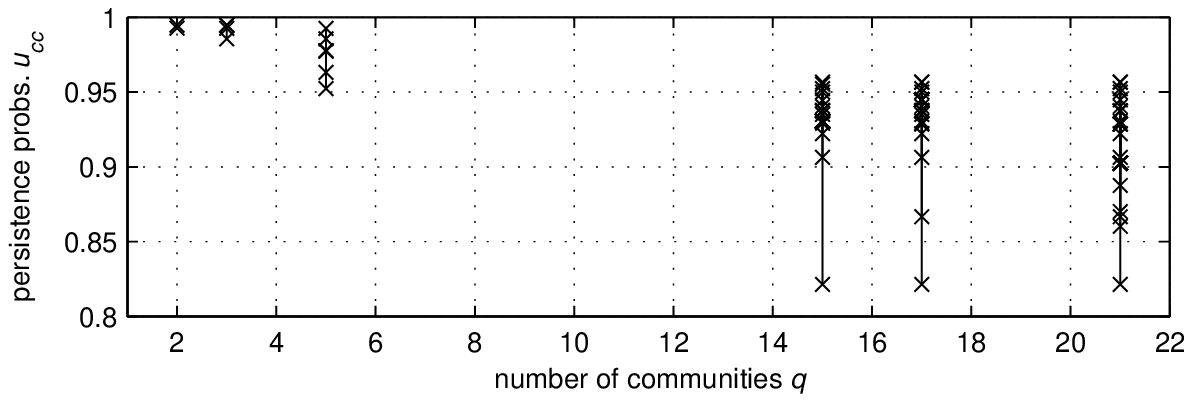}
    \caption{Comparison of two persistence probabilities' diagrams for the Netscience network (the two plots are in the same scale). Above: blow-up of the diagram of Fig. \ref{fig:netscience_ucc} (our results). Below: the diagram related to the six partitions proposed in \cite{DeYa:10}. }\label{fig:netscience_ucc_comp}
    \end{center}
\end{figure}

It is instructive to compare these results with those obtained, on the same case study, by the \textit{graph stability} approach proposed by Delvenne et al. \cite{DeYa:10} (a detailed comparison of the two methods is in the next section). By means of the KVV algorithm \cite{KaVe:04} (a hierarchical, divisive, non-binary, graph clustering method), they obtain a sequence of six partitions, with $q=2,3,5,15,17,21$. Analyzing and comparing the \textit{stability curve} (i.e., the autocovariance function of a signal emitted by a random walker) of each of them, the authors suggest their partition with $q=5$ as the more reliable, as it has the largest stability over a longer time span with respect to any other.
We created the persistence probabilities' diagram of the six partitions of \cite{DeYa:10}, and compared it with our diagram in Fig. \ref{fig:netscience_ucc_comp}. The partition $q=5$ of \cite{DeYa:10} confirms to be definitely more significant than those with finer decomposition (i.e., $q=15,17,21$) according to our criterion too. Actually, our and their $\mathbb{P}_5$ partitions share the same minimal $u_{cc}=0.952$, due to a common $22$-node community. The two partitions are, however, partially different (the normalized mutual information is $I=0.886$, with about $6\%$ of differently classified node pairs).

The inspection of Fig. \ref{fig:netscience_ucc_comp} also reveals that, for each given $q$, the partitions obtained with our method are superior than those proposed in \cite{DeYa:10}, provided the criterion put forward in this paper is adopted. In fact, they are $\alpha$-partitions with an $\alpha$ value which is larger (or at least not smaller) in all six cases. Actually, while the criterion of \cite{DeYa:10} ranks partitions by ``averaging'' among the communities, our approach is a ``worst-case'' one: by selecting an $\alpha$-partition one guarantees that the ``worst'' community has a persistence probability not less than $\alpha$. Finally, note that in the gap from $q=6$ to $15$, where no partition is obtained by the KVV divisive algorithm, our partition generating algorithm provide a set of finer and finer partitions, whose quality only slowly deteriorates as $q$ increases. The analyst of the network can fruitfully select in this interval a proper trade-off between fine granularity and significance of the partition.

\subsection{\label{sec:neural}Neural network}

\begin{figure}
    \begin{center}
    \includegraphics[width=8cm]{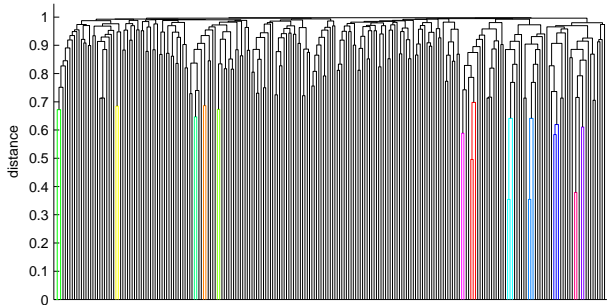}
    \caption{Neural network. The dendrogram obtained with $T=3$. } \label{fig:neural}
    \end{center}
\end{figure}

The third example concerns a directed, weighted network, representing the neural connections of the worm \textit{Caenorhabditis elegans}. Starting from Watts and Strogatz's seminal work \cite{WaSt:98}, different versions of this graph have become a standard benchmark for network analysis. We consider the directed, weighted version (whose largest connected component has $N=239$ nodes), which does not display a definite community structure. In fact, the maximum modularity (estimated as in \cite{BlGu:08}) is rather small, namely $Q=0.486$, if compared to other examples of comparable dimension (e.g., $Q=0.831$ for the Netscience network). The less clusterized structure emerges even visually from the dendrogram of Fig. \ref{fig:neural}, where only few groups of nodes appear well separated from the rest (compare, e.g., with Fig. \ref{fig:lfr}).

\begin{figure}
    \begin{center}
    \includegraphics[width=8cm]{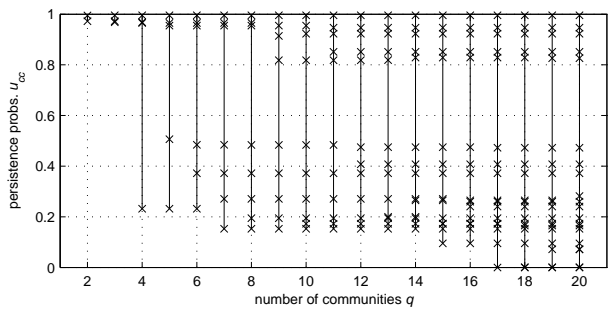}
    \caption{The persistence probabilities' diagram of the neural network. }\label{fig:neural_ucc}
    \end{center}
\end{figure}

We show that our method is able to detect such groups, namely to isolate well-defined communities even in a network which overall does not possess a definite clusterized structure. Consider the persistence probabilities' diagram of Fig. \ref{fig:neural_ucc}. With the exception of the trivial cases $q=2$ and $3$, no $\alpha$-partition exists with $\alpha$ reasonably large. Nonetheless, a few $\alpha$-communities with $\alpha\geq 0.8$ appear and are stably detected in a rather wide range of $q$. More precisely, the same set of five communities with $u_{cc}\geq 0.826$ are revealed in the range $14\leq q\leq 20$. They are clusters, of dimension ranging from $18$ to $29$ nodes, with comparatively rather strong internal connectivity. Any other candidate cluster, instead, turns out to have a much smaller $u_{cc}$ value and, therefore, it cannot be considered to be a significant community.

%%%%%%%%%%%%%%%%%%%%%%%%%%%%%%%%%%%%%%%%%%%%%%%%%%%%%%%

\section{\label{sec:conclusions}Discussion and Conclusions}

In this paper, we have shown that associating a lumped Markov chain to a given network partition (i.e., a set of communities) provides an effective tool for testing the significance of each single community and, consequently, of the entire partition. As a matter of fact, the diagonal terms (called persistence probabilities) of the lumped Markov matrix can be used as  quality measures for each individual community. If a threshold level $0<\alpha<1$ is fixed, a sharp criterion for defining a community as ``significant'' is therefore that of requiring that its persistence probability is not less than $\alpha$.

If an effective method for generating a set of ``good'' partitions is available, the above criterion can be used to rapidly select one of them among those complying with the prescribed $\alpha$-significance, typically the one with the finest network decomposition (i.e., the largest number of communities). We have used a generator of partitions based on hierarchical cluster analysis, where the node distance is again defined on the basis of a Markov chain random walk model. Overall, the method has fair computational requirements, and can be applied to fully general networks (i.e., directed and weighted). Its effectiveness has been demonstrated on several medium-scale examples.

The proposed approach has important connections with two recently published community analysis methods. Delvenne et al. \cite{DeYa:10} show that the autocorrelation function of a signal emitted by a random walker, with value $c$ as long as the walker is in a node $i\in\mathbb{C}_c$, can be expressed in terms of the clustered autocovariance matrix $R_t=H'\left[\mathrm{diag}\left(\pi\right)P^t-\pi'\pi\right]H$,
and they define the stability of the partition $H$ as         $r_t^H=\min_{s=0,1,\ldots,t}\mathrm{trace}\left(R_s\right)$.
Given a set of candidate partitions, the \textit{graph stability} function   $r_t=\max_H r_t^H$ puts in evidence, for each time instant $t$, which is the ``optimal'' partition. It is suggested in \cite {DeYa:10} that the most relevant partitions are those which are optimal over long time windows. It is straightforward to check that our matrix $U$ is related to the step-$1$ autocovariance $R_1$ by $R_1+\Pi'\Pi=\mathrm{diag}(\Pi)U$. The two methods are thus based on the same ground, but our approach has two advantages: first, for each partition $H$ we do not compute a long time-dependent sequence $R_1,R_2,\ldots,R_{t_{\max}}$ (with $t_{\max}$ of the same order as $N$) of $q\times q$ matrices, but the sole matrix $U$, with an important reduction in the computational burden. Second, the full list of the persistence probabilities $u_{cc}$ allows one to test the significance of each single community, whereas the stability of the clustering $r_t^H$ averages among all the communities.

Another work with important connections is that of Weinan et al. \cite{WeLi:08}, who suggest to explicitly find the ``best'' (in a suitable sense) $q$-state approximated lumped Markov chain. This boils out to the formulation of a minimization problem, after a metric on the space of stochastic matrices is introduced. A drawback of this method is however that $q$ must be \textit{a priori} specified, whereas often identifying the correct number of communities is the main goal of the analysis. For the same reason, it can hardly support the discussion of the significance and convenience of choosing one partition instead of another. We argue that this method could be used, jointly with the one proposed in this paper, to generate a set of partitions with increasing values of $q=2,3,\ldots$, by repeatedly solving the above problem. Then, their significance could be tested with the tool of the persistence probabilities' diagram. It is not guaranteed, however, that the proposed partitions are ``good'' in terms of the minimal $u_{cc}$ (i.e., that they are $\alpha$-partitions with large $\alpha$). It is therefore a point deserving further study.

\end{article}
%%%%%%%%%%%%%%%%%%%%%%%%%%%%%%%%%%%%%%%%%%%%%%%%%%%%%%%%%%%%%%%%

%% Adding Figure and Table References
%% Be sure to add figures and tables after \end{article}
%% and before \end{document}

%% For figures, put the caption below the illustration.
%%
%% \begin{figure}[h]
%% \caption{Almost Sharp Front}\label{afoto}
%% \end{figure}

%% For Tables, put caption above table
%%
%% Table caption should start with a capital letter, continue with lower case
%% and not have a period at the end
%% Using @{\vrule height ?? depth ?? width0pt} in the tabular preamble will
%% keep that much space between every line in the table.

%% \begin{table}
%% \caption{Repeat length of longer allele by age of onset class}
%% \begin{tabular}{@{\vrule height 10.5pt depth4pt  width0pt}lrcccc}
%% table text
%% \end{tabular}
%% \end{table}

%% For two column figures and tables, use the following:

%% \begin{figure*}
%% \caption{Almost Sharp Front}\label{afoto}
%% \end{figure*}

%% \begin{table*}
%% \caption{Repeat length of longer allele by age of onset class}
%% \begin{tabular}{ccc}
%% table text
%% \end{tabular}
%% \end{table*}

%%%%%%%%%%%%%%%%%%%%%%%%%%%%%%%%%%%%%%%%%

\end{document}